\documentclass[12pt]{article}
\usepackage{amsmath}
\usepackage[mathcal]{euscript}
\usepackage{graphicx}
\def \myfigures #1#2#3#4#5#6#7#8
{\begin{figure}[p]
    \begin{center}
        \includegraphics[width=#2 \textwidth, angle=-90]{#1.eps}
        \hfill
        \includegraphics[width=#6 \textwidth, angle=-90]{#5.eps}
        \parbox[t]{#4\textwidth}{\caption {#3}\label{#1}}
        \hfill
        \parbox[t]{#8\textwidth}{\caption {#7}\label{#5}}
    \end{center}
\end{figure} }

\begin{document}
\title{Bianchi type-I string cosmological model in the presence of a
magnetic field: classical versus loop quantum cosmology approaches}

\author{Victor Rikhvitsky, Bijan Saha
\thanks{E-mail:~~~bijan@jinr.ru; URL: http://bijansaha.narod.ru}\\
{\small \it Laboratory of Information Technologies}\\
{\small \it Joint Institute for Nuclear Research, Dubna}\\
{\small \it 141980 Dubna, Moscow region, Russia}
\and
Mihai Visinescu
\thanks{E-mail:~~~mvisin@theory.nipne.ro}\\
{\small \it Department of Theoretical Physics}\\
{\small \it Horia Hulubei National Institute for Physics and Nuclear
Engineering}\\
{\small \it  P.O.B. MG-6, 077125 Magurele, Romania}
}
\date{}

\maketitle

\begin{abstract}

A Bianchi type-I  cosmological model in the presence of a magnetic
flux along a cosmological string is considered. The first objective
of this study is to investigate Einstein equations using a tractable
assumption usually accepted in the literature. Quantum effects of the
present cosmological model are examined in the framework of loop
quantum cosmology. Finally we draw a parallel between the classical and
quantum approaches.

~

Pacs: 95.30.Sf; 98.80.Jk; 04.60.Pp

Key words:  Bianchi type-I  model, cosmological
string, magnetic field, loop quantum cosmology
\end{abstract}

\section{Introduction}

The Bianchi models, which describe homogeneous but anisotropic
spacetimes, have been extensively discussed in the literature,
motivated in part by attempts to explain small but significant
anisotropies in the cosmic microwave background (CMB) \cite{JD,EK}
and large structures \cite{MT}.

Among the anisotropic Bianchi models, the simplest ones are Bianchi
type-I (BI) models whose spatial sections are flat but the expansion
or contraction rates are direction dependent.

Primordial magnetic fields can have a significant impact on the CMB
anisotropy. Also the presence of strong magnetic fields raises
interesting problems like the formation of galaxies in the Universe.
The BI models are appropriate for the investigation of a Universe
which is permeated by a large scale, homogeneous magnetic field.

In the early stages of the evolution of the Universe it is expected
that topological defects could  have formed naturally during phase
transitions followed by spontaneous broken symmetries. Cosmic
strings are linear topological defects, have very interesting
properties and might play an important role in structure formation
\cite{VS,HK}.

In the first part of the paper we shall investigate the evolution of a
BI model in presence of a cloud of strings and magnetic field. In order
to solve Einstein equations we resort to a tractable assumption
concerning a relation between the rest energy and  tension
density of the system of strings \cite{PSL}.

Loop Quantum Gravity (LQG) \cite{CR,TT} represents one of the most
compelling attempt towards a complete non perturbative quantum theory
for the gravitational interaction. The cosmological application of LQG
was developed in terms of invariant connections \cite{MB1} and this
model was denoted by Loop Quantum Cosmology (LQC). LQC takes the
ingredients of LQG and applies them to expanding Universe or black
hole models. LQC permits the exploration of the effects of quantum
physics and quantum geometry in gravitation \cite{AA,MB2}.

In order to test the robustness of the LQC it is necessary to apply the
methodology to some concrete situations and one of the most favorable
model is represented by the simplest of anisotropic models, namely BI
cosmologies.
The detailed formulation for LQC in the BI models
\cite{DWC1,CV,DWC2,AE,MV} reveals the fact that gravity can behave
repulsively at Planckian energy densities leading to the replacement of
the big bang singularity with a big bounce.

The plan of the paper is as follows: In Section 2 we outline the
classical equations for a BI string cosmological model in the presence
of a magnetic field. In Section 3 we describe the effective loop
dynamics for the present BI cosmological model. In the next section we
present some numerical simulations and compare the classical and LQC
approaches. Some conclusions and open problems are discussed in the
last Section.

\section{Classical equations}

The line element of a BI Universe is
\begin{equation}
ds^2 = - dt^2 + a_1^2 dx^2 + a_2^2 dy^2 + a_3^2 dz^2\,, \label{BI}
\end{equation}
with three scale factors $a_i$ $(i=1,2,3)$ which are functions
of time $t$ only and consequently three expansion
rates. In principle all these scale factors could be different and it
is useful to express the mean expansion rate in terms of the average
Hubble rate:
\begin{equation}\label{Hubble}
H = \frac{1}{3}(H_1 + H_2 + H_3) = \frac{1}{3}\Bigl(\frac{\dot
a_1}{a_1}+\frac{\dot a_2}{a_2}+ \frac{\dot a_3}{a_3}\Bigr) =
\frac{1}{3} \frac{\dot V}{V}\,,
\end{equation}
where we have defined a new function
\begin{equation}
V = \sqrt{-g}= a_1 a_2 a_3 \,, \label{taudef}
\end{equation}
which is in fact the volume scale of the BI space-time. $H_i$ are
the so-called directional Hubble parameters:
\begin{equation}
H_i = \frac{\dot a_i}{a_i}. \label{DH}
\end{equation}
In \eqref{Hubble}, \eqref{DH}  and further over-dot means
differentiation with respect to $t$.

The Einstein's gravitational field equation has the form
\begin{subequations}
\label{BID}
\begin{eqnarray}
\frac{\ddot a_2}{a_2} +\frac{\ddot a_3}{a_3} +
\frac{\dot a_2}{a_2}\frac{\dot
a_3}{a_3}&=& - \kappa T_{1}^{1}\,,\label{11}\\
\frac{\ddot a_3}{a_3} +\frac{\ddot a_1}{a_1} +
\frac{\dot a_3}{a_3}\frac{\dot
a_1}{a_1}&=&  -\kappa T_{2}^{2}\,,\label{22}\\
\frac{\ddot a_1}{a_1} +\frac{\ddot a_2}{a_2} +
\frac{\dot a_1}{a_1}\frac{\dot
a_2}{a_2}&=& - \kappa T_{3}^{3}\,,\label{33}\\
\frac{\dot a_1}{a_1}\frac{\dot a_2}{a_2} +\frac{\dot
a_2}{a_2}\frac{\dot a_3}{a_3}+\frac{\dot a_3}{a_3}\frac{\dot
a_1}{a_1}&=& - \kappa T_{0}^{0}\,, \label{00}
\end{eqnarray}
\end{subequations}
where $\kappa$ is the gravitational constant.

The energy momentum tensor for a system of cosmic strings and
magnetic field in a comoving coordinate is given by
\begin{equation}
T_{\mu}^{\nu} =  \rho_{string} u_\mu u^\nu - \lambda x_\mu x^\nu +
E_\mu^\nu\,, \label{imperfl}
\end{equation}
where $\rho_{string}$ is the rest energy density of strings with
massive particles attached to them and can be expressed as
$\rho_{string} = \rho_{p} + \lambda$, where $\rho_{p}$ is the rest
energy density of the particles attached to the strings and
$\lambda$ is the tension density of the system of strings
\cite{PSL,pradhan,tade} which may be positive or negative. Here
$u_i$ is the four velocity and $x_i$ is the direction of the string,
obeying the relations
\begin{equation}
u_iu^i = -x_ix^i = -1, \quad u_i x^i = 0\,. \label{velocity}
\end{equation}

In \eqref{imperfl} $E_{\mu\nu}$ is the electromagnetic field given by
Lichnerowich \cite{lich}. In our case the electromagnetic field tensor
$F_{\mu \nu}$ has only one non-vanishing component, namely
\begin{equation}
F_{23} = h\,,
\label{f23}
\end{equation}
where $h$ is assumed to be constant.
For the electromagnetic field $E_\mu^\nu$ one gets the following non-trivial
components
\begin{equation}
E_0^0 = E_1^1 = - E_2^2 = - E_3^3 = \frac{h^2}
{2 {\bar \mu} a_2^2 a_3^2}
\equiv \frac{1}{2}\frac{\beta^2}{(a_2 a_3)^2}\,,
\label{E}
\end{equation}
where $\bar \mu$ is a constant characteristic of the medium and called
the magnetic permeability. Typically $\bar \mu$ differs from unity
only by a few parts in $10^5$ ($\bar \mu > 1$ for paramagnetic
substances and $\bar \mu < 1$ for diamagnetic).

Choosing the string along $x^1$ direction and using comoving
coordinates we have the following components of energy momentum
tensor \cite{ass}:
\begin{eqnarray}
T_{0}^{0} + \rho_{string} =  T_{1}^{1} + \lambda = - T_{2}^{2} = -
T_{3}^{3} = \frac{\beta^2}{2}\frac{a_1^2}{V^2}\,. \label{total}
\end{eqnarray}

Taking into account the conservation of the energy-momentum tensor,
i.e., $T_{\mu;\nu}^{\nu} = 0$, after a little manipulation of
\eqref{total} one obtains \cite{SV,SRV}:
\begin{equation}\label{rholambda}
\dot \rho_{string} + \frac{\dot V}{V}\rho_{string} - \frac{\dot
a_1}{a_1}\lambda = 0\,.
\end{equation}
Here we take into account that the conservation law for magnetic
field fulfills identically.

It is customary to assume a relation between $\rho_{string}$ and
$\lambda$ in accordance with the state equations for strings. The
simplest one is a proportionality relation \cite{PSL}:
\begin{equation}\label{rhoalphalambda}
\rho_{string} = \alpha \lambda \,.
\end{equation}
The  most usual choices of the constant $\alpha$ are
\begin{equation}\label{alpha}
\alpha =\left \{
\begin{array}{ll}
1 & \quad {\rm geometric\,\,\,string}\\
1 + \omega  & \quad \omega \ge 0, \quad p \,\,{\rm string\,\,\,or\,\,\,
Takabayasi\,\,\,string}\\
-1  & \quad {\rm Reddy\,\,\,string}\,.
\end{array}
\right.
\end{equation}

From eq. \eqref{rholambda} with \eqref{rhoalphalambda} we get
\begin{equation}\label{rhostring}
\rho_{string} = R a_1^{\frac{1-\alpha}{\alpha}} a_2^{-1} a_3^{-1}\,,
\end{equation}
with $R$ a constant of integration.

Let us also write the other features of BI metric such as expansion
and shear. The expansion for the BI metric takes the form
\begin{equation}
\vartheta = \frac{\dot a_1}{a_1}+\frac{\dot a_2}{a_2}+ \frac{\dot
a_3}{a_3} = \frac{\dot V}{V}, \label{expan}
\end{equation}
while the nonzero components for the shear tensor read
\begin{equation}
\sigma_i \equiv \sigma_i^i = \frac{\dot a_i}{a_i} -
\frac{1}{3}\vartheta.
\label{sigcom}
\end{equation}
In \eqref{sigcom} and henceforth there is no summation over repeated
index "$i$". The shear energy density in given by
\begin{eqnarray}\label{shed}
\Sigma^2 = \frac{1}{2}\sigma_{\mu\nu}\sigma^{\mu\nu} = 
\frac{1}{6}\left( (H_1-H_2)^2 + (H_2-H_3)^2 + (H_3-H_1)^2 \right)\,.
\end{eqnarray}

\section{Effective loop quantum dynamics}
In the loop quantum cosmology approach we shall use a Hamiltonian
framework where the degrees of freedom of the Bianchi type-I model are
encoded in the triad components $p_i$ and momentum components $c_i$
as follows:
\begin{equation}\label{pc}
p_1 = a_2 a_3, \quad  p_2 = a_1 a_3, \quad  p_3 = a_1 a_2, \quad c_i
= \gamma \dot{a_i} \,.
\end{equation}
Here $\gamma$ is the Barbero-Immirzi parameter and represents a quantum
ambiguity of loop quantum gravity which is a non-negative real valued
parameter.

In terms of these variables, the total Hamiltonian of the model is
\begin{eqnarray}\label{htot}
\mathcal{H} &=&   \mathcal{H}_{grav} + \mathcal{H}_{matter} \nonumber\\
&=&  \frac{-1}{\kappa \gamma^2  \sqrt{p_1 p_2 p_3} }
(c_2 c_3 p_2 p_3 + c_1 c_3 p_1 p_3 + c_1 c_2 p_1 p_2)
+ \sqrt{p_1 p_2 p_3} \rho_M \,,
\end{eqnarray}
where $\rho_M$ is the matter energy density. In our model $\rho_M$
comprises the contribution of cosmological string density
$\rho_{string}$ given by \eqref{rhostring}:
\begin{equation}
\rho_{string} = R p_1^{- \frac{\alpha + 1}{2\alpha}} (p_2
p_3)^{\frac{1 - \alpha}{2 \alpha}}, \label{rhostringlqc}
\end{equation}
and the energy density of the magnetic field \eqref{E} \cite{MV}
\begin{equation}
\rho_{mag} = \frac{1}{2} \frac{\beta^2}{(a_2 a_3)^2} = \frac{1}{2}
\frac{\beta^2}{p_1^2}\,. \label{rhomaglqc}
\end{equation}

Einstein's equations are derived from Hamilton's equations:
\begin{equation}\label{ham}
\dot{p_i}= \{p_i,\mathcal{H}\} =  -\kappa \gamma \frac{\partial
\mathcal{H}}{\partial c_i}, \quad  \dot{c_i}= \{c_i,\mathcal{H}\} =
\kappa \gamma \frac{\partial \mathcal{H}}{\partial p_i}\,.
\end{equation}
On the other hand, the total Hamiltonian $\mathcal{H}$ is of constrained type whereby it
vanishes identically for solutions of Einstein's equations
\begin{equation}\label{h0}
\mathcal{H} = 0 \,.
\end{equation}

Using the explicit form of the Hamiltonian $\mathcal{H}$ we have for
$p_1$ the following equation
\begin{equation}\label{p1}
\frac{d p_1}{d t} = \frac{p_1}{\gamma \sqrt{p_1 p_2 p_3} }
(c_2 p_2 + c_3 p _3)\,,
\end{equation}
and similar equations for $p_2$ and $p_3$. For the evolution of $c_i$
we get:
\begin{subequations}
\label{ci}
\begin{eqnarray}
\frac{d c_1}{d t} =&-&\frac{c_1}{\gamma \sqrt{p_1 p_2 p_3} }
(c_2 p_2 + c_3 p _3)\nonumber\\
&+& \frac{1}{2 \gamma p_1  \sqrt{p_1 p_2 p_3} }
(c_2 c_3 p_2 p_3 + c_1 c_3 p_1 p_3 + c_1 c_2 p_1 p_2) \nonumber\\
&+& \frac{\kappa \gamma}{p_1} \left[-\frac{1}{2\alpha} R
\left(\frac{p_2 p_3}{p_1}\right)^{\frac{1}{2\alpha}}
-\frac{3}{4} \beta^2 \left(\frac{p_2 p_3}{p_1^3}\right)^{\frac{1}{2}}
\right] \,,\label{c1}
\end{eqnarray}
\begin{eqnarray}
\frac{d c_2}{d t} =&-&\frac{c_2}{\gamma  \sqrt{p_1 p_2 p_3} }
(c_1 p_1 + c_3 p _3)\nonumber\\
&+& \frac{1}{2 \gamma p_2  \sqrt{p_1 p_2 p_3} }
(c_2 c_3 p_2 p_3 + c_1 c_3 p_1 p_3 + c_1 c_2 p_1 p_2) \nonumber\\
&+& \frac{\kappa \gamma}{p_2} \left[\frac{1}{2\alpha} R
\left(\frac{p_2 p_3}{p_1}\right)^{\frac{1}{2\alpha}} +
\frac{1}{4} \beta^2 \left(\frac{p_2 p_3}{p_1^3}\right)^{\frac{1}{2}}
\right] \,,\label{c2}
\end{eqnarray}
\begin{eqnarray}
\frac{d c_3}{d t} =&-&\frac{c_3}{\gamma  \sqrt{p_1 p_2 p_3} }
(c_1 p_1 + c_2 p _2)\nonumber\\
&+& \frac{1}{2 \gamma p_3  \sqrt{p_1 p_2 p_3} }
(c_2 c_3 p_2 p_3 + c_1 c_3 p_1 p_3 + c_1 c_2 p_1 p_2) \nonumber\\
&+& \frac{\kappa \gamma}{p_3} \left[\frac{1}{2\alpha} R
\left(\frac{p_2 p_3}{p_1}\right)^{\frac{1}{2\alpha}}  +
\frac{1}{4} \beta^2 \left(\frac{p_2 p_3}{p_1^3}\right)^{\frac{1}{2}}
\right] \,.\label{c3}
\end{eqnarray}
\end{subequations}

Let us observe that from equations for $p_i$ and $c_i$ we have
the following relation:
\begin{equation}
\frac{d}{d t} (p_i c_i) = \kappa \gamma \sqrt{p_1 p_2 p_3}
\left (\frac{1}{2} \rho_M + p_i \frac{\partial
\rho_M}{\partial p_i}\right )\,.
\end{equation}
The directional Hubble rates now reads
\begin{equation}
H_i = \frac{\dot{a_i}}{a_i} =  \frac{\sqrt{p_i} c_i}{\gamma
\sqrt{p_j p_k}}\,, \quad i \ne j \ne k = 1,\,2,\,3. \label{DHlqc}
\end{equation}
Hamiltonian \eqref{htot} on account of the vanishing condition
\eqref{h0} and \eqref{DHlqc} leads to
\begin{equation}
H_1 H_2 + H_1 H_3 + H_2 H_3 = \kappa \rho_M,
\end{equation}
which is in fact the zero-zero component \eqref{00} of the Einstein
system of equations \eqref{BID}.

Taking into account the symmetry of the density $\rho_M$ with respect
to variables $p_2$ and $p_3$ we have
\begin{equation}\label{23}
\frac{d}{dt} (p_2 c_2 - p_3 c_3) = 0\,.
\end{equation}
This means that for the directional Hubble parameters $H_2\,,H_3$ we
have
\begin{equation}\label{H23}
H_2 - H_3 = \frac {\alpha_{23}}{a_1 a_2 a_3} = \frac {\alpha_{23}}
{\sqrt{p_1 p_2 p_3}}\,,
\end{equation}
with $\alpha_{23}$ a constant.

The quantum effects in loop quantum cosmology arise in the {\it
effective} Hamiltonian constructed from the classical one by replacing
the classical $c_i$ terms with sine functions:
\begin{equation}\label{csin}
c_i \longrightarrow \frac{\sin (\bar \mu_i' c_i)}{\bar \mu_i'}\,.
\end{equation}
where $\bar \mu_i$ are real valued functions of the triad coefficients
$p_i$.

The effective Hamiltonian is given by:
\begin{equation}\label{heff}
\mathcal{H}_{eff} =
\frac{-1}{\kappa \gamma^2 \sqrt{p_1 p_2 p_3}}
\left\{\frac{\sin ({\bar\mu_2'} c_2)
\sin ({\bar \mu_3'}c_3)}{{\bar\mu_2'}{\bar \mu_3'}}  p_2 p_3 +
\text{cyclic~terms} \right \} + \sqrt{p_1 p_2 p_3} \rho_M \,,
\end{equation}
and the Hamilton's equations for $p_i$ and $c_i$  will be modified
accordingly.

It is quite evident that in the limit $\bar \mu_i \rightarrow 0$, the
classical Hamiltonian $\mathcal{H}$ \eqref{htot} is recovered.
The expression of the  parameters $\bar \mu_i'$ as functions of the
triad components $p_i$ represent an ambiguity of the quantization. Two
most preferable constructions are discussed in \cite{CV,DWC2,AE}.

In what follows we shall adopt the $\bar \mu'$-scheme in which the
parameters $\bar \mu_i'$ are chosen as follows:
\begin{equation}
\bar \mu_1' = \sqrt{\frac{p_1 \Delta}{p_2 p_3}}\,,
\bar \mu_2' = \sqrt{\frac{p_2 \Delta}{p_1 p_3}}\,,
\bar \mu_3' = \sqrt{\frac{p_3 \Delta}{p_1 p_2}}\,,
\end{equation}
with $\Delta$ a constant related to the minimum area gap in LQG. For
the numerical simulations it is assumed that $\Delta = O(1)$ in Planck
units.

Let us remark that from the vanishing of the Hamiltonian \eqref{h0} we
have the bound:
\begin{equation}\label{bound}
p_1 p_2 p_3 \rho_M \leq \frac{1}{\kappa \gamma^2}
\left\{\frac{p_2 p_3}{\bar\mu_2 \bar\mu_3} +
\frac{p_1 p_3}{\bar\mu_1 \bar\mu_3}+
\frac{p_1 p_2}{\bar\mu_1 \bar\mu_2} \right\}\,.
\end{equation}
In particular, in the $\bar\mu'$ scheme the total density is bounded
by the critical value:
\begin{equation}\label{rhocrit}
\rho_{M\,crit} = 3(\kappa \gamma^2 \Delta)^{-1}\,,
\end{equation}
implying that the classical singularity is never approached. Indeed the
total energy of the matter must be below this value and the classical
collapse is replaced by a bounce.

\section{Numerical simulations and a comparison of the approaches}

The complexity of the equations does not allow for analytic solutions
and imposes numerical simulations. The classical equations of motion
given by \eqref{p1}-\eqref{ci} and the corresponding ones for quantum
effects in LQC approach with the replacement \eqref{csin} can be solved
once the initial values $p_i (t = t_0)$ and $c_i (t = t_0)$ are given.

In what follows we report some numerical studies on the behavior of
$\rho_M$, $V$  and the anisotropic
shear $\sigma_{\mu\nu}$. Taking into account that only diagonal
components of shear tensor are non-zero, in the new variables they
now read
\begin{equation}\label{shear}
\sigma_i =
\frac{\sqrt{p_i} c_i}{\gamma \sqrt{p_j p_k}} -
\frac{1}{3\gamma}\Bigl(\frac{\sqrt{p_1} c_1}{\sqrt{p_2 p_3}} +
\frac{\sqrt{p_2} c_2}{\sqrt{p_3 p_1}} + \frac{\sqrt{p_3}
c_3}{\sqrt{p_1 p_2}}\Bigr)\,, \quad i \ne j \ne k = 1,\,2,\,3\,,
\end{equation}
and the shear energy density is
\begin{eqnarray}\label{shedlqc}
\Sigma^2 = \frac{1}{6 \gamma^2 p_1 p_2 p_3}\bigl[(p_1c_1 - p_2c_2)^2
+ (p_2c_2 - p_3c_3)^2 + (p_3c_3 - p_1c_1)^2\bigr]\,.
\end{eqnarray}

In doing so we considered a number of cases that helps us to clarify
the role of various parameters. To begin with we examined both
positive and negative $\alpha$. In particular we considered the case
with $\alpha = 2$ and $\alpha = -2$ and it was found that the value or
sign of $\alpha$ leaves the overall picture qualitatively unchanged.
As a second consideration we set a large value of $R$ and small
value of $\beta$ and vice versa, for example, $R =
18.24,\,\beta = 1$ and $R = 0.90,\, \beta = 6.5$, respectively .
It was established that for both cases the overall picture remains
qualitatively unaltered. Finally we consider the case setting
different initial values for $c_i = c_0$.

In Figs. \ref{fig1}, \ref{fig3}, \ref{fig5}, \ref{fig7} we have
illustrated the evolution of volume scale $V$ (red solid line),
energy density $\rho_M$ (blue dash line) and shear energy $\Sigma^2$
(black dot line), whereas in Figs. \ref{fig2}, \ref{fig4},
\ref{fig6}, \ref{fig8} we have plotted the evolution of the
components of shear tenor $\sigma_1$ (red solid line), $\sigma_2$
(blue dash line) and  $\sigma_3$ (black dot line).

In the classical case the initial condition $c_0 <0$ leads to a
collapsing Universe [cf. Figs. \ref{fig1}, \ref{fig2}], while
$c_0 > 0$ gives rise to an expanding one [cf. Figs. \ref{fig3},
\ref{fig4}].

In the LQC approach, even for $c_0 <0$ the energy density remains
bounded below the critical energy \eqref{rhocrit} as expected from
analytical considerations [cf. Fig. \ref{fig5}]. After the bounce, the
regime is an expanding one and the Universe isotropizes [cf. Fig.
\ref{fig6}]. On the other hand, for $c_0 > 0$ we have only expanding
phase of the Universe, the shear energy density $\Sigma^2$ remains
finite [cf. Fig. \ref{fig7}] and again $\sigma_i$ tend to zero
[cf. Fig. \ref{fig8}].


\myfigures{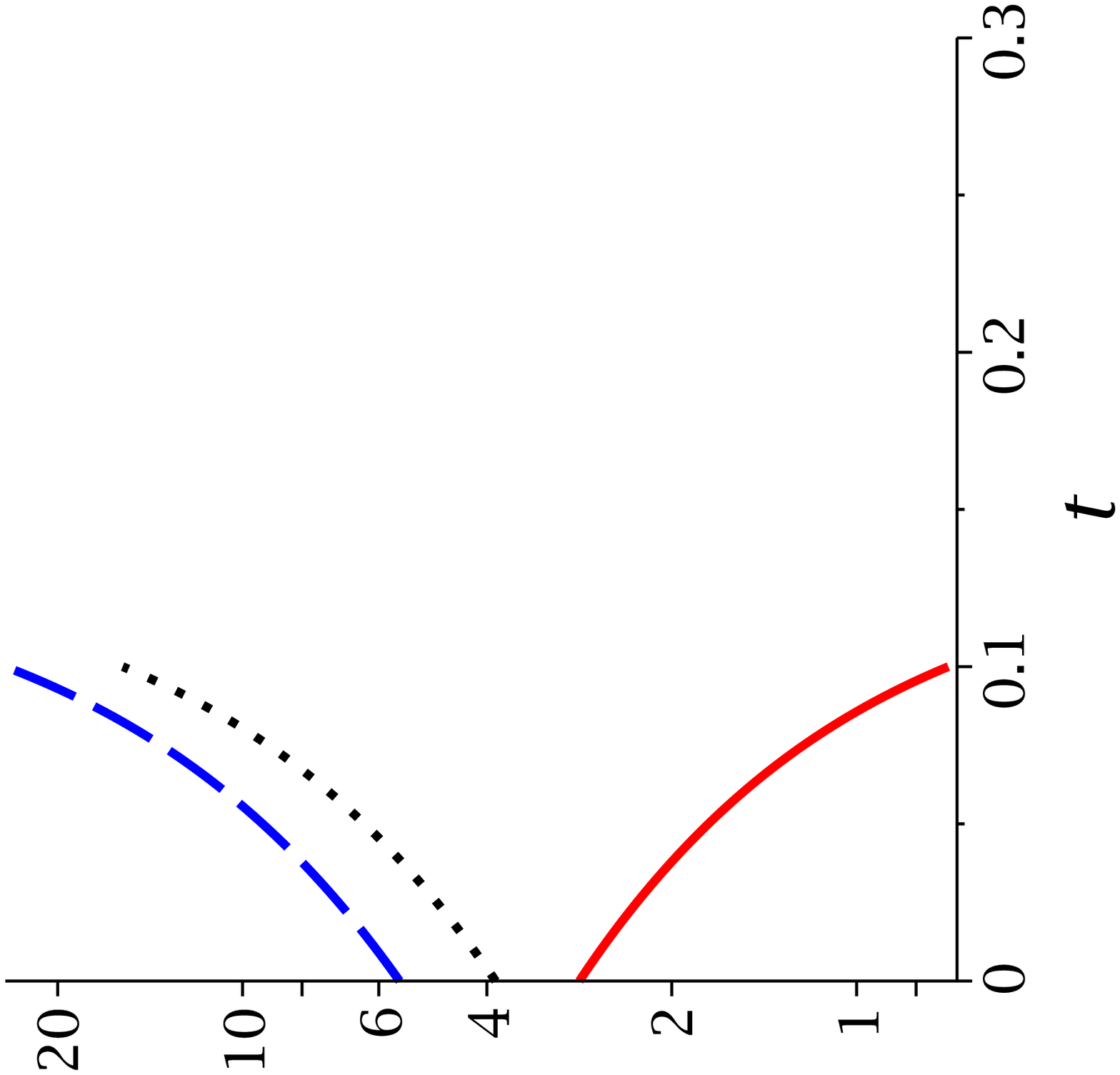}{0.4} {Classical: Evolution of $V$, $\rho_M$ and
$\Sigma^2$ for $\alpha=-2$, $R=18.24$, $\beta=1$, $c_0 = -1$. Here
and further the red solid line corresponds to $V$, blue dash line
to $\rho_M$ and black dot line  to $\Sigma^2$.}{0.4}
{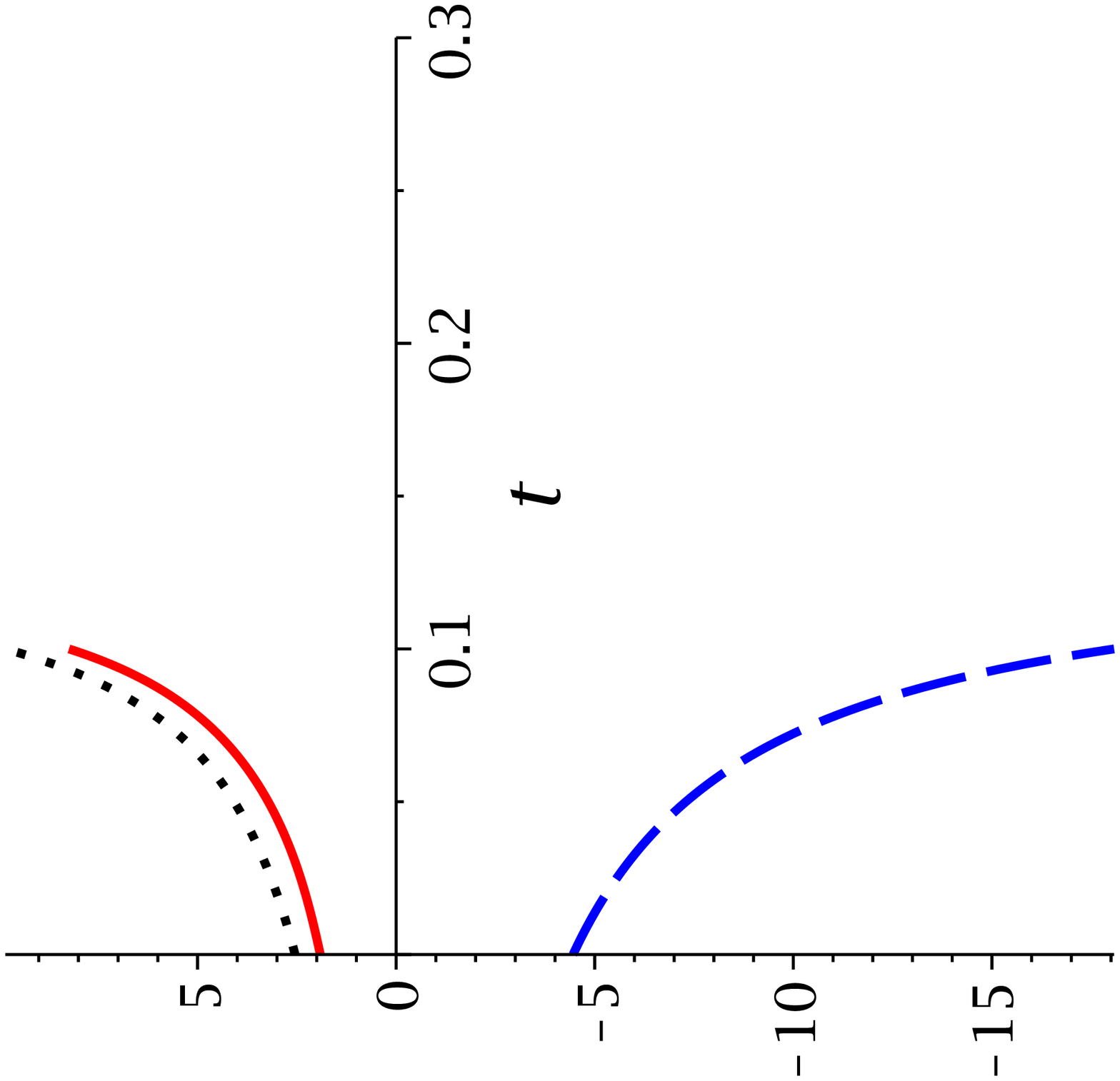}{0.4} {Classical: Evolution of $\sigma_i$'s
for $\alpha=-2$, $R=18.24$, $\beta=1$, $c_0 = -1$.  Here and further
the red solid line corresponds to $\sigma_1$, blue dash line
to $\sigma_2$ and black dot line to $\sigma_3$.}{0.4}


\myfigures{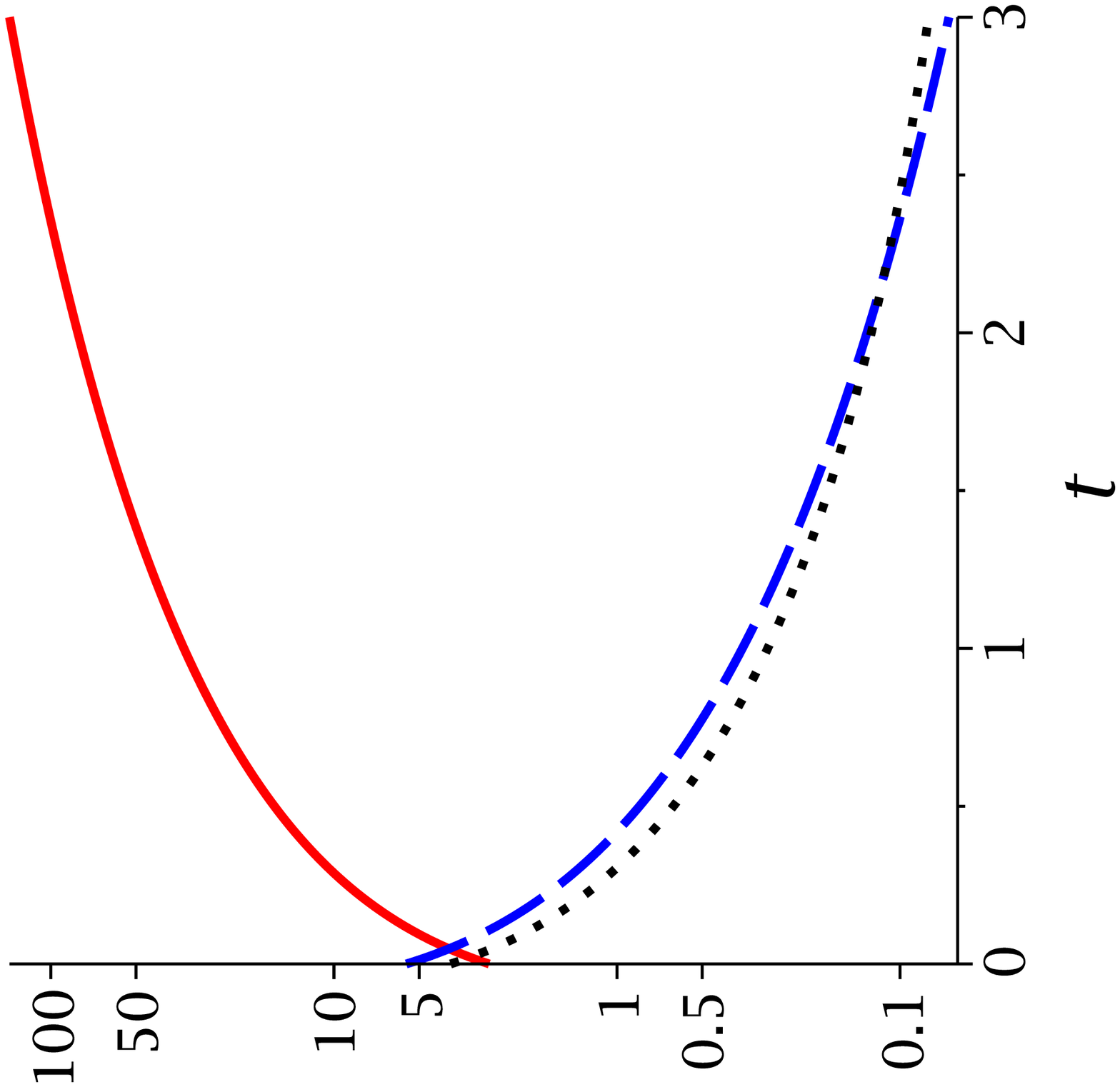}{0.4} {Classical: Evolution of $V$, $\rho_M$ and
$\Sigma^2$ for $\alpha=-2$, $R=18.24$, $\beta=1$, $c_0 = 1$.}{0.4}
{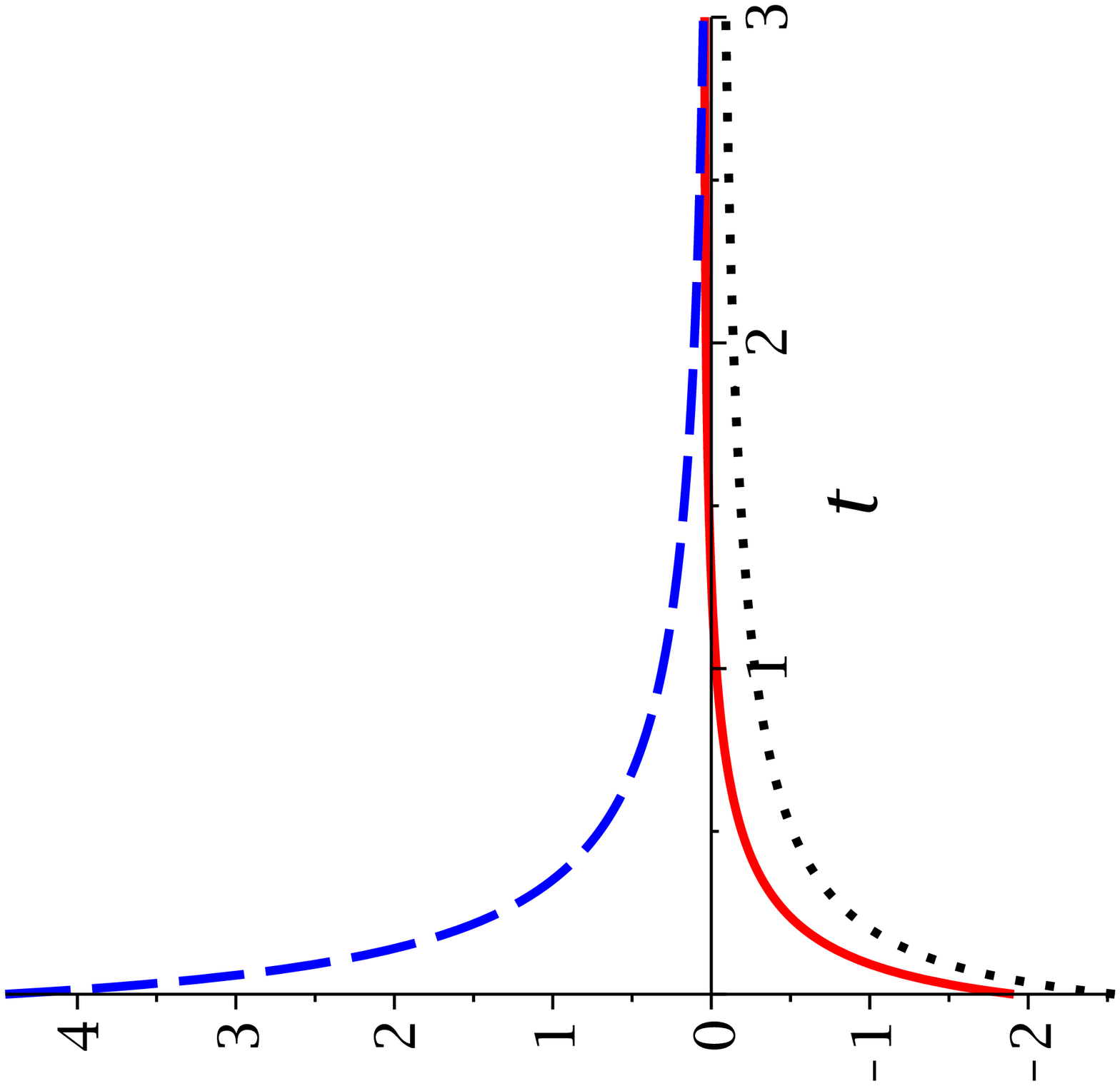}{0.4} {Classical: Evolution of $\sigma_i$'s for $\alpha=-2$,
$R=18.24$, $\beta=1$, $c_0 = 1$.}{0.4}


\myfigures{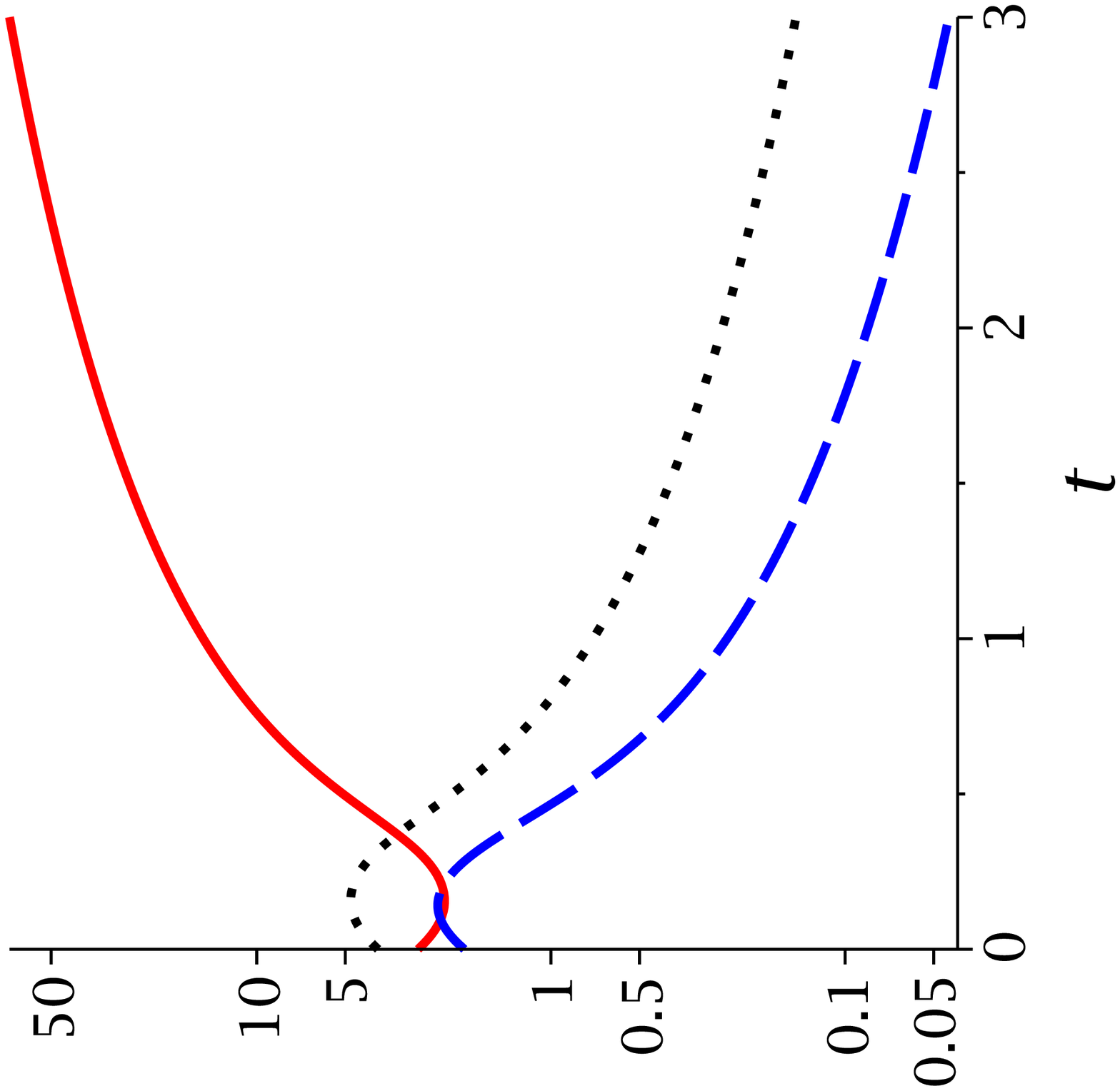}{0.4} {LQC: Evolution of $V$, $\rho_M$ and
$\Sigma^2$ for $\alpha=-2$, $R=6.20$, $\beta=1$, $c_0 = -1$.}{0.4}
{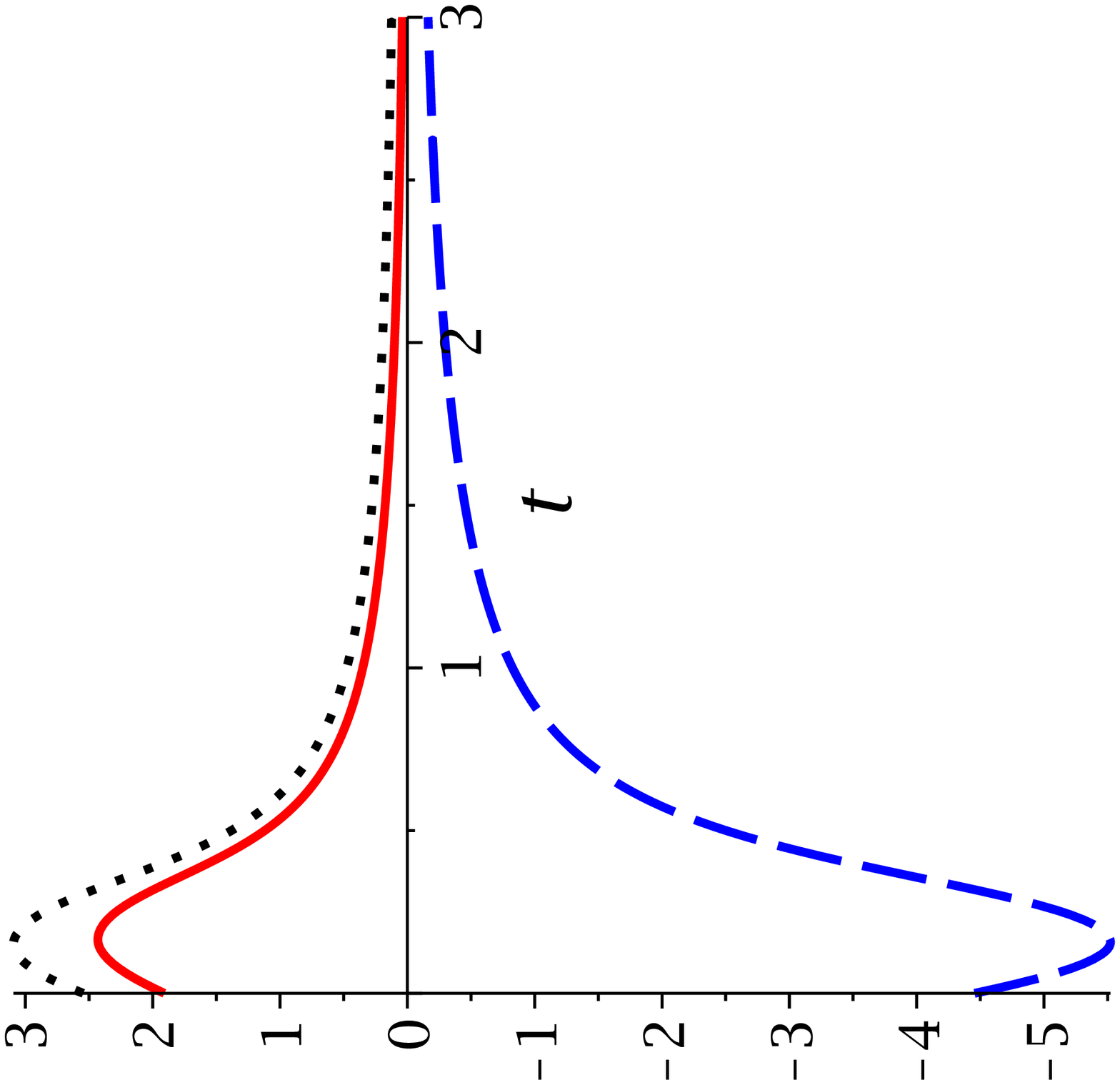}{0.4} {LQC: Evolution of $\sigma_i$'s for $\alpha=-2$,
$R=6.20$, $\beta=1$, $c_0 = -1$.}{0.4}


\myfigures{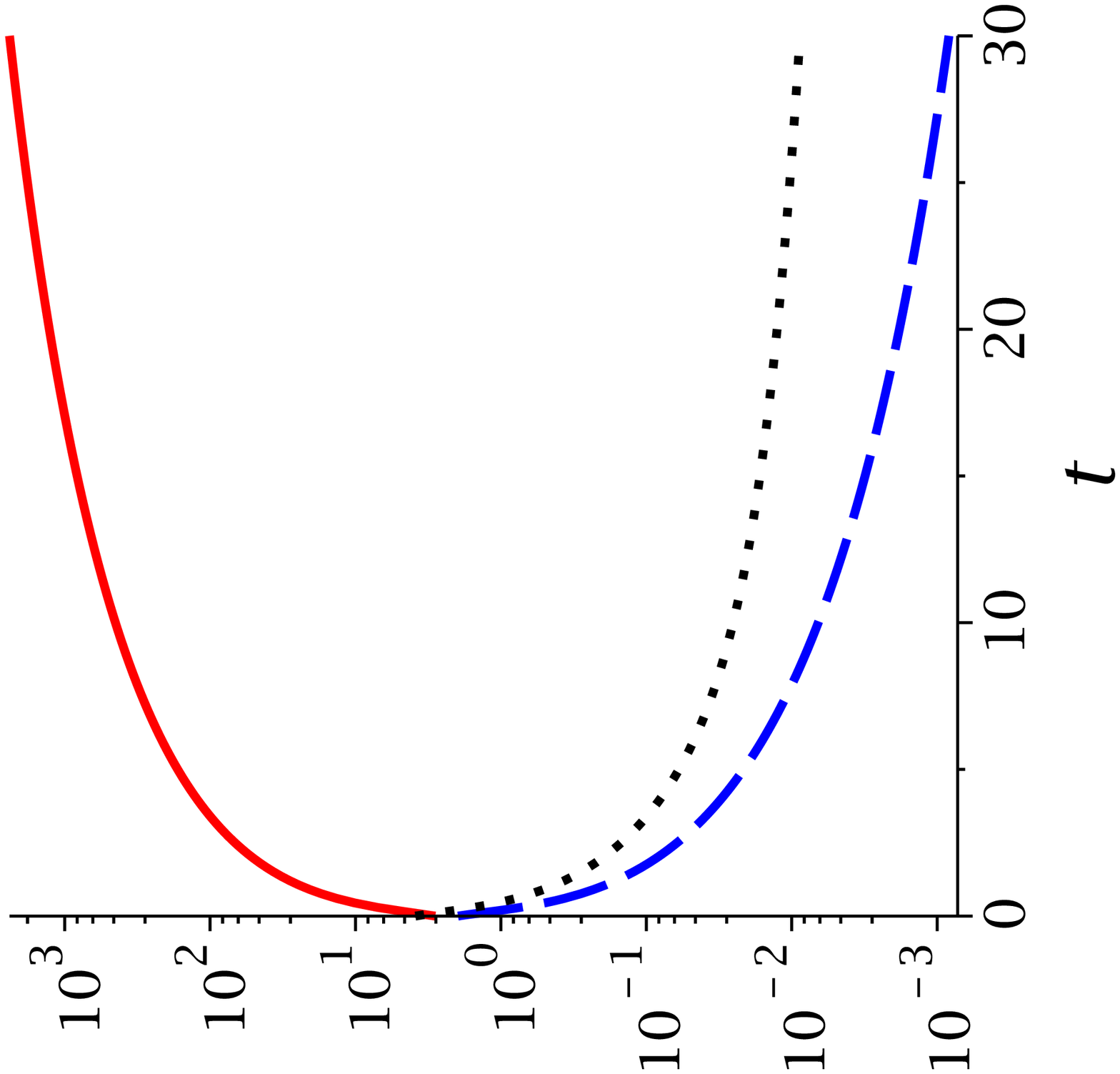}{0.4} {LQC: Evolution of $V$, $\rho_M$ and
$\Sigma^2$ for $\alpha=-2$, $R=R=6.20$, $\beta=1$, $c_0 = 1$.}{0.4}
{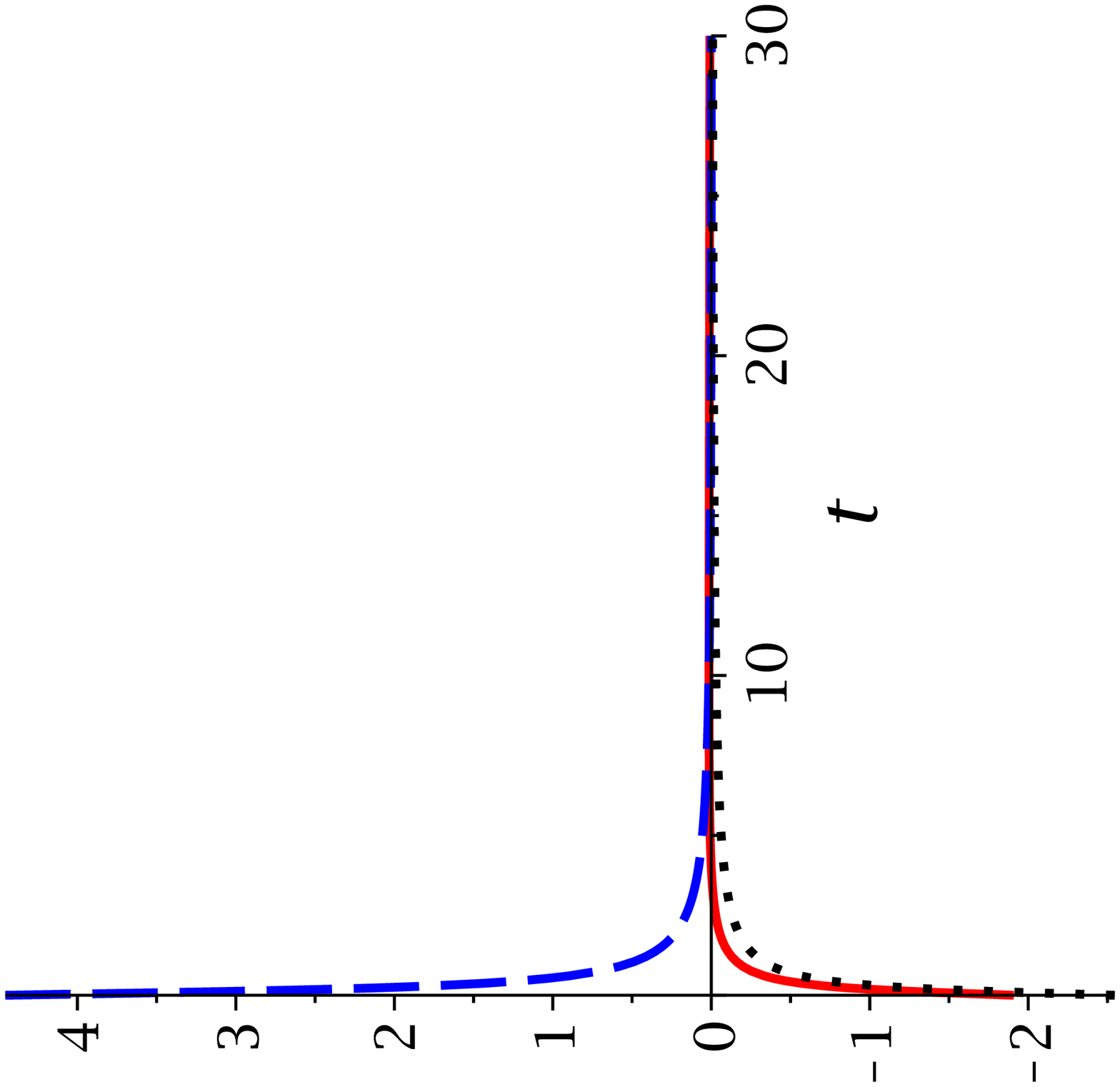}{0.4} {LQC: Evolution of $\sigma_i$'s for $\alpha=-2$,
$R=6.20$, $\beta=1$, $c_0 = 1$.}{0.4}


\section{Conclusions}

In this report within the scope of Bianchi type-I cosmological model
we investigate the role of cosmic string and magnetic field on the
evolution of the Universe. In doing so we employ both classical and
LQC approaches. It is found that the qualitative picture of
evolution does not depend on the cosmic string ($\alpha$), though
the value of $\alpha$ leads to quantitative changes. On the other
hand magnetic field together with cosmic string (given by the pair
$\beta$ and $R$) also leave the qualitative feature unaltered. Only
initial value of $c_i$'s, i.e., initial rate of change of the metric
functions $a_i$'s play essential role.

In the classical approach the initial condition $c_0 < 0$ corresponds
to a classically collapsing Universe, while $c_0 > 0$ is associated
with an expansion. In the LQC approach for $c_0 < 0$ the singularity is
avoided via a bounce. After the bounce the Universe enters an expansion
phase with an asymptotic isotropization. For positive $c_0$ we get
always expansion and isotropization.

At the classical level other Bianchi models have richer phenomenology
than the cosmological BI model. From this point of view the extension of
the string cosmological model in the presence of electromagnetic fields
to other types of anisotropies and comparisons between the classical
approach and LQC one deserve further studies.

\subsection*{Acknowledgments}

This work is supported in part by a joint Romanian-LIT, JINR, Dubna
Research Project, theme no. 05-6-1060-2005/2013. M.V. is partially
supported by program  PN-II-ID-PCE-2011-3-0137, Romania.


%

\begin{thebibliography}{99}
%
\bibitem{JD}
J. Donkley {\it et al.} [WMAP Collaboration],
{\em Astrophys. J. Suppl.} {\bf 180}, 306 (2009).
%
\bibitem{EK}
E. Komatsu {\it et al.} [WMAP Collaboration],
{\em Astrophys. J. Suppl.} {\bf 192}, 17 (2011).
%
\bibitem{MT}
M. Tegmark {\it et al.} [SDSS Collaboration],
{\em Phys. Rev. D} {\bf 69}, 103501 (2004).
%
\bibitem{VS}
A. Vilenkin, E. P. S. Shellards,
{\it Cosmic Strings and Other Topological Defects},
Cambridge University Press, Cambridge (1994).
%
\bibitem{HK}
M. B. Hindmarsh, T. W. B. Kible,
{\em Rep. Prog. Phys.} {\bf 58}, 477 (1995).
%
\bibitem{PSL}
P. S. Letelier, {\it Phys. Rev. D} {\bf 28}, 2414
(1983).
%
\bibitem{CR}
C. Rovelli,
{\it Quantum Gravity},
Cambridge University Press, Cambridge (2004).
%
\bibitem{TT}
T. Thiemann,
{\it Modern Canonical Quantum General Relativity},
Cambridge University Press, Cambridge (2006).
%
\bibitem{MB1}
M. Bojowald,
{\em Class. Quantum Grav.} {\bf 17}, 1489 (2000).
%
\bibitem{AA}
A. Ashtekar,
{\em Gen. Relativ. Grav.} {\bf 41}, 707 (2009).
%
\bibitem{MB2}
M. Bojowald,
{\em Class. Quantum Grav.} {\bf 26}, 075020 (2009).
%
\bibitem{DWC1}
D.-W. Chiou,
{\em Phys. Rev. D} {\bf 75}, 024029 (2007).
%
\bibitem{CV}
D.-W. Chiou and K. Vandersloot,
{\em Phys. Rev. D} {\bf 76}, 084015 (2007).
%
\bibitem{DWC2}
D. W. Chiou,
{\em Phys. Rev. D} {\bf 76}, 124037 (2007).
%
\bibitem{AE}
A. Ashtekar and E. Wilson-Ewing,
{\em Phys. Rev. D} {\bf 80}, 123532 (2009).
%
\bibitem{MV}
R. Maartens and K. Vandersloot,
{\em arXiv: 0812.1889 [gr-qc]}.
%
\bibitem{pradhan}
A. Pradhan, A. K. Yadav, R. P. Singh, V. K. Singh,
{\it Astrophys. Space Sci.} {\bf 312}, 145 (2007).
%
\bibitem{tade}
G.S. Khadekar and S.D. Tade,
{\it Astrophys. Space Sci.} {\bf 310}, 47 (2007).
%
\bibitem{lich} A. Lichnerowicz,
{\it Relativistic Hydrodynamics and Magnetohydrodynamics},
Benjamin, New York (1967).
%
\bibitem{ass}
B. Saha, {\it Astrophys. Space Sci.} {\bf 299}, 149 (2005).
%
\bibitem{SV}
B. Saha and M. Visinescu, {\it Astrophys. Space Sci.} {\bf
315}, 99 (2008).
\bibitem{SRV}
B. Saha, V. Rikhvitsky and M. Visinescu, {\it Cent. Eur. J.
Phys.} {\bf 8}, 113 (2010).
%
%
\end{thebibliography}
\end{document}